**Measuring the effect of Non-Pharmaceutical Interventions (NPIs) on mobility during the COVID-19 pandemic using global mobility data**

Berber T Snoeijer[1], MSc; Mariska Burger[2], B.Com.Hons; Shaoxiong Sun[3], PhD; Richard JB Dobson[3,4], PhD; Amos A Folarin*[3,4], PhD

[1]ClinLine, Leiderdorp, The Netherlands
[2]OCS Life Sciences, 's Hertogenbosch, The Netherlands
[3]The Department of Biostatistics and Health informatics, Institute of Psychiatry, Psychology and Neuroscience, King's College London, London, UK
[4]Institute of Health Informatics, University College London, London, UK

**\*Corresponding author**
Amos A. Folarin
King's College London
SGDP Centre, IoPPN
PO Box 80
De Crespigny Park, Denmark Hill
London SE5 8AF
Email: amos.folarin@kcl.ac.uk

# Abstract

The implementation of governmental Non-Pharmaceutical Interventions (NPIs) has been the primary means of controlling the spread of the COVID-19 disease. The intended effect of these NPIs has been to reduce mobility. A strong reduction in mobility is believed to have a positive effect on the reduction of COVID-19 transmission by limiting the opportunity for the virus to spread in the population. Due to the huge costs of implementing these NPIs, it is essential to have a good understanding of their efficacy. Using global mobility data, released by Apple and Google, and ACAPS NPI data, we have investigated the proportional contribution of NPIs on i) the size of the change (magnitude) of transition between pre- and post-lockdown mobility levels and ii) the rate (gradient) of this transition. Using generalized linear models to find the best fit model, we found similar results using Apple or Google data. NPIs found to have impacted the magnitude of the change in mobility were: lockdown measures (Apple, Google Retail and Recreation (RAR) and Google Transit and Stations (TS)); declaring a state of emergency (Apple, Google RAR and Google TS); closure of businesses and public services (Google RAR) and school closures (Apple). Using cluster analyses and chi square tests, we found that closure of businesses and public services, school closures and limiting public gatherings, as well as border closures and international flight suspensions, were closely related. The implementation of lockdown measures and limiting public gatherings have had the greatest effect on the rate of mobility change. In conclusion, we were able to quantitatively assess the efficacy of NPIs in reducing mobility, which have enabled us to understand their fine grained effects in a timely manner, thereby facilitating well-informed and cost-effective interventions.

# Introduction

There is little question that COVID-19 has had a huge impact on global health and economies, with costs of $11 Trillion, projected (as of June 2020) by the International Monetary Fund (IMF) [1]. The implementation of governmental Non-Pharmaceutical Interventions (NPIs) has been the primary means of controlling the pandemic and range in type and implementation. The key intended effect of these NPIs, such as school closure and limitations of public gatherings, has been the constraint of mobility. Mobility refers to the unrestricted movement of the population within a country. A strong reduction in mobility is believed to have a positive effect on the reduction of COVID-19 transmission by limiting the opportunity for the virus to spread in the

population, and so reducing the "R0" effective reproduction number - which indicates to what extent each infected individual infects new uninfected individuals.

Although there is probably much variation in how NPIs were actually executed in each individual country, an overall effect in reduction of the COVID-19 cases and mortality has been observed [2-5]. However, due to the huge costs of implementing these NPIs, we reason that it is essential to have a good understanding of their benefits. The variation in effectiveness can as well be proxied from the extent in which mobility is reduced in each country, derived from freely available aggregated phone-derived mobility data issued by Apple and Google [6,7]. Apple mobility data reflects requests for directions in Apple Maps for driving, walking and train transit, while Google mobility data shows movement trends by country across different categories of places, like retail and recreation (RAR), grocery and pharmacy, transit and stations (TS), parks, workplaces and residential. The effects of constrained mobility on disease reduction, as well as on the world economy, are needed in a timely manner and at a local level as countries/regions emerge from these measures and need to rapidly identify sub-populations of cases. Smartphones provide a pre-existing, location aware, passive means to observe mobility at both population and individual level, in real time.

Deriving meaningful standardized measures from real world data, like Google and Apple mobility rates, is challenging. Weekday effects, holiday effects and unexpected events effects vary from period to period and from country to country and are reflected in the data. Standardized modelling techniques might be less useful for this kind of real world time series data. Considering mobility and NPI data from several countries, this study aims to establish a useful model from which the proportional effects of the NPIs on the state transition between pre- and post-lockdown can be assessed. Two standardized measures were calculated: i) the rate of transition between the states (gradient) and ii) the magnitude of the change [Figure 1]. These measures are also usable for global studies on COVID-19 spread and economic effects, as they depend on existing data and could practically be implemented in a useful timeframe.

# Methods

## Data Sources & Availability

All data sources used in this study were freely accessible and available from their respective providers.

ACAPS global dataset of NPIs

The Assessment Capacities Project (ACAPS) has collected a Governmental Measures dataset in which every action taken by governments in response to COVID-19 falling into their taxonomy is captured [8]. The NPIs are reported and captured on a country level across five categories: governance and socio-economic measures, lockdown, movement restrictions, public health measures and social distancing. Within each category there are several different NPIs [Table 1]. NPIs were further classified as being targeted to a specific subpopulation or on a national level. As we assessed the overall country population effect, we disregarded the NPIs that were targeted at a specific subpopulation. NPIs could be re-issued or prolonged, which is recorded in the ACAPs dataset. For the purpose of the analysis, only the first implementation date for each NPI was used. NPI data extracted on 14 June 2020 was used in the analysis.

**Table 1. All Non-Pharmaceutical NPIs (NPIs)**

| Category | Non-Pharmaceutical Intervention (NPI) |
|---|---|
| Governance and socio-economic measures | Economic measures |
| | Emergency administrative structures activated or established |
| | Limit product imports/exports |
| | Military deployment |
| | State of emergency declared |
| Lockdown | Full lockdown |
| | Lockdown of refugee/idp camps or other minorities |
| | Partial lockdown |
| Movement restrictions | Additional health/documents requirements upon arrival |
| | Border checks |
| | Border closure |
| | Checkpoints within the country |
| | Complete border closure |
| | Curfews |
| | Domestic travel restrictions |
| | International flights suspension |
| | Surveillance and monitoring |
| | Visa restrictions |
| Public health measures | Amendments to funeral and burial regulations |
| | Awareness campaigns |
| | General recommendations |
| | Health screenings in airports and border crossings |
| | Isolation and quarantine policies |
| | Mass population testing |
| | Other public health measures enforced |
| | Psychological assistance and medical social work |
| | Requirement to wear protective gear in public |
| | Strengthening the public health system |
| | Testing policy |
| Social distancing | Changes in prison-related policies |
| | Closure of businesses and public services |
| | Limit public gatherings |
| | Schools closure |

Apple mobility direction data

The Apple mobility data reflects requests for directions in Apple Maps for driving, walking and train transit [6]. The data shows a relative volume of direction requests per country compared to a baseline volume on 13 January 2020. In order to make the data comparable with the Google mobility data, the percentage change values were added or subtracted from 100%. For the purpose of reflecting overall mobility, we calculated the average from the driving, walking and train transit data for each country. Apple data was extracted on 14 June 2020 and was used in the analysis for the time period from 13 January 2020 to 14 June 2020.

Google mobility data

The Google data shows movement trends by country across different categories of places, like retail and recreation (RAR), grocery and pharmacy, transit and stations (TS), parks, workplaces and residential [7]. The data is based on users who turned on the Google mobility tracking history and shows how the relative amount of visitors to categorized places have changed, compared to baseline. Baseline was defined as the median value from the corresponding weekday in the 5-week period from 3 January 2020 to 6 Feb 2020. By using the baseline of the corresponding weekday, the weekday effect is reduced. Google data was extracted on 7 June 2020 and was used for the time period of 15 February 2020 to 7 June 2020. Only the TS and RAR data have been analysed as a clear lockdown effect was only visible in these Google data sources.

## Analysis methods and derivations

All analyses were performed using SAS® Studio 3.7 Enterprise Edition. As the Apple data was less variable, the primary focus was on the Apple data. The Google data was used for comparison and validation.

To assess the effect of NPIs on the magnitude and speed of the reduction of mobility, the following two primary parameters were derived:
1. The average % mobility lost after stabilizing to assess the magnitude of the effect of the NPIs.
2. The gradient of line connecting the start of the decline and the stabilizing point to assess the speed of the effect of the NPIs.

Time series data during lockdown was smoothed to account for the weekday effect by adding the average weekday deviation to the mobility. As the weekday effect was not consistent for the pre- and post-lockdown period, this was only performed for post-lockdown by starting from the first Sunday where the mobility decreased below 70%. The weekday deviation was calculated as the difference between the actual mobility on the specific weekday and the average mobility for that corresponding week.

The initial cut-off value for start of stable mobility was defined as the midpoint between 100% and the average of the 5 lowest reported mobility rates. The initial ***stable mobility date*** was then calculated as the first date where the mobility dropped below this initial cut-off value and did not increase more than 10% (absolute day-to-day change) for at least 7 consecutive days. The ***average % remaining mobility after stabilizing*** was calculated as the average of the mobility for the 2 weeks following the stable mobility date and then subtracted from 100% to result in the ***average % mobility lost after stabilizing***. To get a more precise estimate of the start of stable mobility, an ***adjusted stable mobility date*** was determined for each country as the first date where the mobility dropped below the average % remaining mobility after stabilizing (excluding any initial decreases followed by a steep increase again). Visual inspection of the resulting parameters showed an accurate estimation for nearly all countries included in both Apple and Google data (see [https://github.com/MariskaBurger/Covid-19-analysis-results/tree/master](https://github.com/MariskaBurger/Covid-19-analysis-results/tree/master)).

In order to calculate the speed of the effect of the NPIs, the ***start of the decline*** of the mobility was calculated as the last occurrence before the adjusted stable mobility date 1) with changes in mobility rate of less than +7% (so slight increases were also considered); 2) without 3 or more consecutive changes between -2% and 7%, and 3) with a total decline from the beginning of the decline to the last declining observation of more than 60% of the average % mobility lost after stabilizing.

This was done to exclude periods where the % mobility stayed relatively constant for a period of time before declining. These cut-off values were determined by visual observation of the data and accounted for potential weekday effects. After finding the start of the decline date for each country, the ***gradient*** of the line connecting the start of the decline and the adjusted stable mobility date was calculated. Figure 1 below presents an example of the Apple mobility data for the Netherlands to visually show the different parameters calculated.

Figure 1. Mobility rate transition in a country pre and post implementation of NPIs: The dotted red line represents the adjusted stable mobility date. The green dotted line represents the start of the decline and the gradient of the solid black line is used to assess the speed of the effect of the NPIs. Lastly the solid blue line shows the average % remaining mobility after stabilizing.

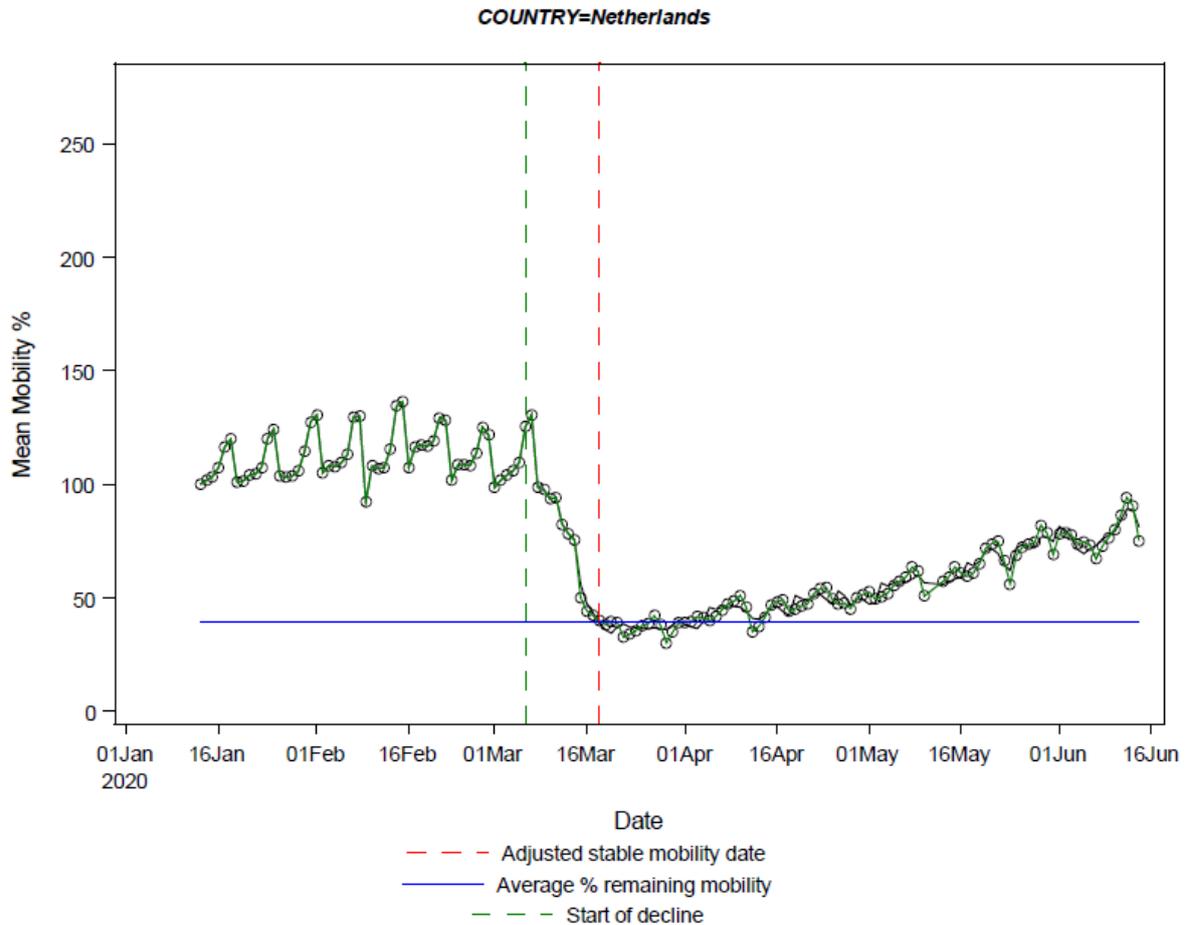

For each country the total **number of unique NPIs** implemented before the adjusted stable mobility date, the number of unique NPIs implemented within 3 weeks before, and the **time between the implementation of the first and last NPI** were calculated.

Cluster analyses were performed to determine which NPIs were closely related in the different countries. Binary values were created for each NPI and country and were assigned the value of 0 if the NPI was not implemented at all, or if the NPI was implemented more than 3 weeks before the adjusted stable mobility date, and the value of 1 was assigned if the NPI was implemented within 3 weeks before the adjusted stable mobility date. Jaccard coefficients between each pair of NPIs were calculated as the number of countries that are coded as 1 for both NPIs (in each pair) divided by the number of countries that are coded as 1 for either or both NPIs (in each pair). The centroid method, measuring the (squared) euclidean distance between the centroids or means of the clusters, was used to determine the unique clusters. The association between the different NPIs was further analysed using a Chi-square test for all possible combinations of NPIs. The same binary values used in the cluster analyses were also used in the chi-square analyses.

The two primary parameters were analysed, using generalized linear models to determine which NPIs had the biggest or fastest effect on the mobility (gradient and % change). The same binary values (0 if NPI was not implemented or implemented more than 3 weeks before the adjusted stable mobility date, 1 if NPI was implemented within 3 weeks before the adjusted stable mobility date) used for the cluster analyses and for the Chi-square analyses were also used to explore which NPIs had the biggest or fastest effect on the mobility. In all cases the lowest Akaike's Information Criterion (AIC) value was used to determine the best fit model. All assumptions related to generalized linear models were verified to hold true (residual errors are independent, normally distributed and have constant variance).

**Code availability**

Code for data cleaning and analysis is provided as part of the replication package. It is available at https://github.com/MariskaBurger/Covid-19-analysis-results/tree/master for review. It will be uploaded to the journal repository once the paper has been conditionally accepted.

# Results

The mobility rate for each country is presented graphically, including the start of the decline date, the adjusted stable mobility date and the average % remaining mobility. The Apple data is presented for 59 countries and the Google data for 124 countries. See visualisations at https://github.com/MariskaBurger/Covid-19-analysis-results/tree/master. Due to variation in the Google data, not all parameters could be calculated for all countries.

For the Apple data, a mean of 8.3 NPIs was implemented within a country before the adjusted stable mobility date, compared to a mean of 6.9 NPIs implemented within the 3 last weeks before the adjusted stable mobility date [Table 2]. The results for the Google data differed slightly from the Apple data, due to slight differences in the assessed adjusted stable mobility data, but were comparable with the Apple data.

Table 2. Summary statistics for the number of NPIs implemented within the countries

| Statistic | Apple (N = 59) | | Google Retail and Recreation (N = 124) | | Google Transit and Stations (N = 124) | |
|---|---|---|---|---|---|---|
| | NPIs before stabilizing | NPIs <= 3 weeks before stabilizing | NPIs before stabilizing | NPIs <= 3 weeks before stabilizing | NPIs before stabilizing | NPIs <= 3 weeks before stabilizing |
| Mean | 8.3 | 6.9 | 8.5 | 7.3 | 8.7 | 7.6 |
| Median | 8.0 | 7.0 | 8.5 | 7.0 | 8.5 | 8.0 |
| Mode | 7 | 7 | 9 | 9 | 8 | 9 |
| Range | 2 - 16 | 1 - 12 | 0 - 17 | 0 - 15 | 0 - 17 | 0 - 14 |

An overview of the timing of the NPIs in relation to the adjusted stable mobility date is presented in Table 3. From this it can be observed that most NPIs were issued within a few days before lockdown. However, NPIs like awareness campaigns and public health measures were often issued earlier than the actual lockdown measures. NPIs that occurred more than 3 weeks before the adjusted stable mobility date were less likely to have an effect on lockdown and seem to have an obscuring effect on the analysis results. By classifying the time before stabilization (as presented in the table below) in the analysis and testing of the resulting models, the 3-week cut-off appeared to give the best results. Therefore we have disregarded NPIs issued before from further analysis. In addition, for the same reason of convergence of models and reliability of model effects, we have excluded interventions that occurred in less than 20% of the countries. The number and percentage of countries which implemented a specific NPI within the last 3 weeks is presented in Table 4. Social distancing measures like the limitation of public gatherings were implemented in the majority of the countries (Apple: 78.0%, Google RAR: 77.4%, Google TS: 77.4%), followed by school closures (Apple: 64.4%,

Google RAR: 69.4%, Google TS: 71.0%) and closure of businesses and public services (Apple: 55.9%, Google RAR: 53.2%).

**Table 3. Percentage of countries implementing NPIs: Period between first NPI implementation and the adjusted stable mobility date for the Apple data**

| Category | NPI | Number of Countries with NPI | Days/Weeks before adjusted stable mobility date | | | | | | | |
|---|---|---|---|---|---|---|---|---|---|---|
| | | | >8 Weeks | 7-8 Weeks | 4-6 Weeks | 2-3 Weeks | 10-14 Days | 7-9 Days | 4-6 Days | 1-3 Days |
| Governance and socio-economic measures | Economic measures | 32 | 3.1 | 3.1 | 9.4 | 3.1 | 3.1 | 12.5 | 31.3 | 34.4 |
| | Emergency administrative structures activated or established | 22 | 9.1 | 18.2 | 9.1 | 4.5 | 4.5 | 18.2 | 18.2 | 18.2 |
| | Limit product imports/exports | 4 | 0.0 | 0.0 | 0.0 | 25.0 | 0.0 | 25.0 | 25.0 | 25.0 |
| | Military deployment | 7 | 0.0 | 0.0 | 0.0 | 0.0 | 14.3 | 0.0 | 14.3 | 71.4 |
| | State of emergency declared | 26 | 0.0 | 3.8 | 0.0 | 3.8 | 7.7 | 15.4 | 42.3 | 26.9 |
| Lockdown | Full lockdown | 2 | 0.0 | 0.0 | 0.0 | 0.0 | 0.0 | 0.0 | 0.0 | 100.0 |
| | Lockdown (Partial/Full) | 18 | 0.0 | 0.0 | 0.0 | 0.0 | 0.0 | 0.0 | 16.7 | 83.3 |
| | Partial lockdown | 17 | 0.0 | 0.0 | 0.0 | 0.0 | 0.0 | 5.9 | 17.6 | 76.5 |
| Movement restrictions | Additional health/documents requirements upon arrival | 8 | 12.5 | 0.0 | 12.5 | 0.0 | 25.0 | 0.0 | 12.5 | 37.5 |
| | Border checks | 11 | 9.1 | 9.1 | 18.2 | 0.0 | 27.3 | 0.0 | 18.2 | 18.2 |
| | Border closure | 27 | 0.0 | 0.0 | 0.0 | 0.0 | 11.1 | 22.2 | 18.5 | 48.1 |
| | Curfews | 1 | 0.0 | 0.0 | 0.0 | 0.0 | 0.0 | 0.0 | 100.0 | 0.0 |
| | Domestic travel restrictions | 10 | 0.0 | 0.0 | 0.0 | 0.0 | 0.0 | 0.0 | 20.0 | 80.0 |
| | International flights suspension | 21 | 4.8 | 0.0 | 0.0 | 0.0 | 4.8 | 14.3 | 33.3 | 42.9 |
| | Surveillance and monitoring | 11 | 18.2 | 9.1 | 18.2 | 0.0 | 0.0 | 9.1 | 27.3 | 18.2 |
| | Visa restrictions | 15 | 0.0 | 0.0 | 0.0 | 0.0 | 20.0 | 20.0 | 26.7 | 33.3 |
| Public health measures | Awareness campaigns | 24 | 8.3 | 25.0 | 8.3 | 8.3 | 8.3 | 4.2 | 16.7 | 20.8 |
| | General recommendations | 28 | 14.3 | 7.1 | 3.6 | 10.7 | 10.7 | 7.1 | 14.3 | 32.1 |
| | Health screenings in airports and border crossings | 22 | 18.2 | 18.2 | 9.1 | 0.0 | 22.7 | 4.5 | 13.6 | 13.6 |
| | Isolation and quarantine policies | 29 | 3.4 | 3.4 | 10.3 | 3.4 | 6.9 | 13.8 | 24.1 | 34.5 |
| | Mass population testing | 3 | 0.0 | 0.0 | 0.0 | 33.3 | 0.0 | 33.3 | 33.3 | 0.0 |
| | Other public health measures enforced | 1 | 0.0 | 0.0 | 0.0 | 100.0 | 0.0 | 0.0 | 0.0 | 0.0 |
| | Psychological assistance and medical social work | 3 | 0.0 | 33.3 | 0.0 | 0.0 | 0.0 | 0.0 | 0.0 | 66.7 |
| | Requirement to wear protective gear in public | 2 | 0.0 | 0.0 | 0.0 | 0.0 | 0.0 | 0.0 | 50.0 | 50.0 |
| | Strengthening the public health system | 33 | 9.1 | 27.3 | 3.0 | 3.0 | 9.1 | 6.1 | 21.2 | 21.2 |
| | Testing policy | 4 | 0.0 | 25.0 | 25.0 | 0.0 | 0.0 | 25.0 | 0.0 | 25.0 |
| Social distancing | Changes in prison-related policies | 8 | 0.0 | 0.0 | 0.0 | 0.0 | 0.0 | 12.5 | 50.0 | 37.5 |
| | Closure of businesses and public services | 33 | 0.0 | 0.0 | 0.0 | 3.0 | 0.0 | 9.1 | 30.3 | 57.6 |
| | Limit public gatherings | 46 | 2.2 | 4.3 | 0.0 | 10.9 | 17.4 | 21.7 | 21.7 | 21.7 |
| | Schools closure | 38 | 0.0 | 0.0 | 2.6 | 0.0 | 7.9 | 28.9 | 28.9 | 31.6 |

**Table 4. Number and percentage of countries in which the NPIs were implemented less than 3 weeks before the adjusted stable mobility date**

| Category | Intervention (NPI) | Apple (N = 59) n (%) | Google Retail and Recreation (N = 124) n (%) | Google Transit and Stations (N = 124) n (%) |
|---|---|---|---|---|
| Governance and socio-economic measures | Economic measures | 32 (54.2) | 53 (42.7) | 55 (44.4) |
| | Emergency administrative structures activated or established | 22 (37.3) | 42 (33.9) | 43 (34.7) |
| | Limit product imports/exports | 4 (6.8)* | 8 (6.5)* | 9 (7.3)* |
| | Military deployment | 7 (11.9)* | 9 (7.3)* | 11 (8.9)* |
| | State of emergency declared | 26 (44.1) | 55 (44.4) | 54 (43.5) |
| Lockdown | Full lockdown | 2 (3.4)* | 4 (3.2)* | 5 (4)* |
| | Lockdown of refugee/IDP camps or other minorities | 0 (0)* | 1 (0.8)* | 1 (0.8)* |
| | Partial lockdown or Full lockdown** | 18 (30.5) | 46 (37.1) | 48 (38.7) |
| | Partial lockdown | 17 (28.8) | 43 (34.7) | 45 (36.3) |
| Movement restrictions | Additional health/documents requirements upon arrival | 8 (13.6)* | 16 (12.9)* | 16 (12.9)* |
| | Border checks | 11 (18.6)* | 14 (11.3)* | 15 (12.1)* |
| | Border closure | 27 (45.8) | 68 (54.8) | 67 (54) |
| | Checkpoints within the country | 1 (1.7)* | 3 (2.4)* | 3 (2.4)* |
| | Complete border closure | 0 (0)* | 1 (0.8)* | 1 (0.8)* |
| | Curfews | 1 (1.7)* | 27 (21.8) | 26 (21) |
| | Domestic travel restrictions | 10 (16.9)* | 32 (25.8) | 36 (29) |
| | International flights suspension | 21 (35.6) | 55 (44.4) | 57 (46) |
| | Surveillance and monitoring | 11 (18.6)* | 20 (16.1)* | 22 (17.7)* |
| | Visa restrictions | 15 (25.4) | 31 (25) | 33 (26.6) |
| Public health measures | Amendments to funeral and burial regulations | 0 (0)* | 7 (5.6)* | 8 (6.5)* |
| | Awareness campaigns | 24 (40.7) | 48 (38.7) | 51 (41.1) |
| | General recommendations | 28 (47.5) | 49 (39.5) | 53 (42.7) |
| | Health screenings in airports and border crossings | 22 (37.3) | 56 (45.2) | 57 (46) |
| | Isolation and quarantine policies | 29 (49.2) | 55 (44.4) | 54 (43.5) |
| | Mass population testing | 3 (5.1)* | 3 (2.4)* | 3 (2.4)* |
| | Other public health measures enforced | 1 (1.7)* | 5 (4)* | 6 (4.8)* |
| | Psychological assistance and medical social work | 3 (5.1)* | 5 (4)* | 4 (3.2)* |
| | Requirement to wear protective gear in public | 2 (3.4)* | 5 (4)* | 7 (5.6)* |
| | Strengthening the public health system | 33 (55.9) | 61 (49.2) | 63 (50.8) |
| | Testing policy | 4 (6.8)* | 9 (7.3)* | 9 (7.3)* |
| Social distancing | Changes in prison-related policies | 8 (13.6)* | 12 (9.7)* | 12 (9.7)* |
| | Closure of businesses and public services | 33 (55.9) | 66 (53.2) | 67 (54) |
| | Limit public gatherings | 46 (78) | 96 (77.4) | 96 (77.4) |
| | Schools closure | 38 (64.4) | 86 (69.4) | 88 (71) |

\* NPIs occurring in <20% of the countries were excluded from the statistical analysis.
\*\* Lockdown NPIs were combined into a single NPI. If a country experienced both partial and full lockdowns the country was only counted once for the combined NPI.

For both Apple and Google data, summary statistics were calculated for the average % mobility lost over all the countries (See Table 5). The average % mobility lost for the Apple data follows a normal distribution with a mean of 60.6%. The smallest % mobility loss was observed for Sweden (30.2%) which did not have a partial or full lockdown before the stabilizing of mobility. The largest % mobility loss was observed for Spain (86.3%) which had a partial lockdown before the stabilization of mobility (see Figure 2 below). Both the Google RAR data as well as the Google TS data also followed a normal distribution with slightly lower mean values, however still comparable with the Apple data.

**Figure 2. Average % mobility lost comparison for Sweden vs Spain**

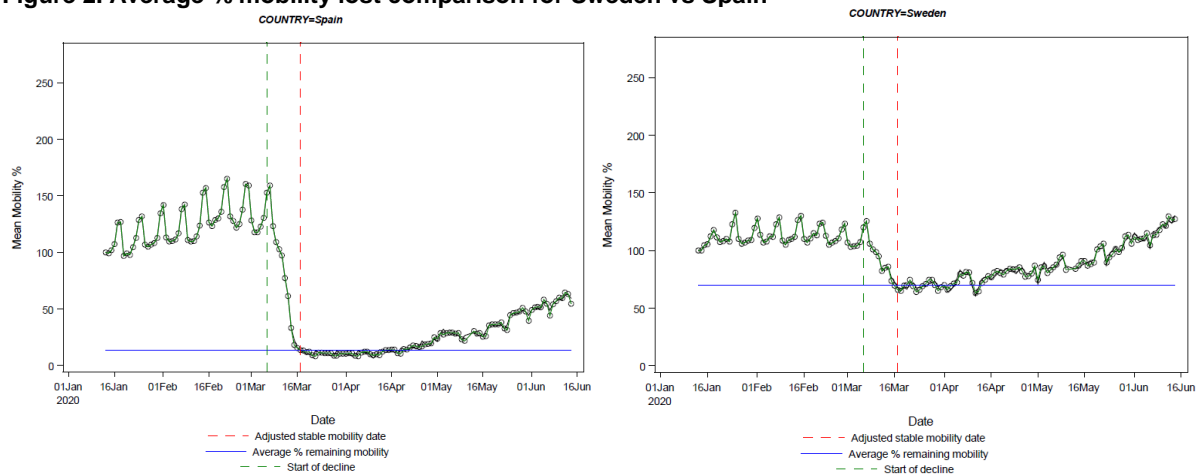

**Table 5. Summary statistics for the average % mobility lost after stabilizing**

| Statistic | Apple (N = 59) | Google Retail and Recreation (N = 124) | Google Transit and Stations (N = 124) |
|---|---|---|---|
| n | 59 | 122 | 123 |
| Mean (SD) | 60.6 (12.20) | 56.9 (17.7) | 57.2 (15.5) |
| Median | 60.3 | 60.3 | 57.5 |
| Range | 30.2 - 86.3 | 19.8 - 89.2 | 20.4 - 86.5 |
| 95% CI | (57.4 ; 63.7) | (53.7 ; 60.1) | (54.4 ; 60.0) |

The speed of the effect of the NPIs, as measured by the gradient of the linear line connecting the start of the decline and the adjusted stable mobility data, was summarized by descriptive statistics (See Table 6). The gradient was calculated for most countries. However, due to the variability for a few countries, it was difficult to establish - as can be seen from the slightly lower number of countries included in these summary statistics. The average rate of reduction in mobility for the Apple data was 7.8% per day and ranged from -27.9 % to -2.9% per day. The country with the fastest effect was South-Africa where the effect of the NPIs was seen almost immediately, while the country with the slowest effect was the United Arab Emirates where the effect of the NPIs was more gradual (See Figure 3).

**Table 6. Summary statistics for the gradient of the line connecting the start of the decline and the adjusted stable mobility date**

| Statistic | Apple (N = 59) | Google Retail and Recreation (N = 124) | Google Transit and Stations (N = 124) |
|---|---|---|---|
| n | 57 | 119 | 118 |
| Mean (SD) | -7.8 (3.6) | -8.7 (12.1) | -6.3 (5.2) |
| Median | -7.1 | -5.5 | -5.1 |
| Range | -27.9 to -2.9 | -96 to -1.13 | -36.5 to -1.1 |
| 95% CI | (-8.7 ; -6.7) | (-10.9 ; -6.5) | (-7.2 ; -5.4) |

SD: Standard Deviation, CI: Confidence Interval

**Figure 3. Gradient comparison for South-Africa versus United Arab Emirates**

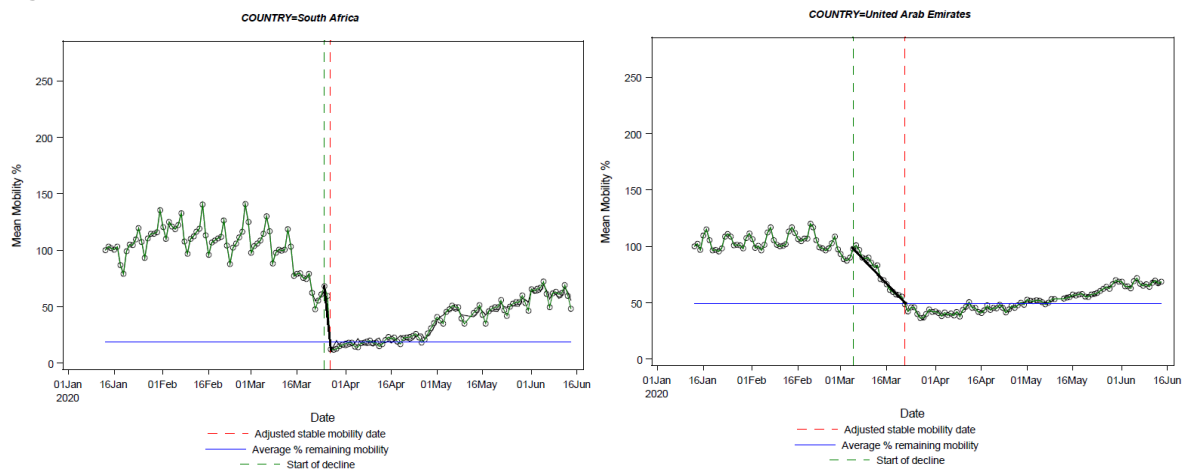

The overall results of the cluster analysis is presented by the Tree diagram in Figure 4. The associations between NPIs like closure of businesses and public services, school closures and limiting public gatherings, as well as border closures and international flight suspensions, are tight - indicating that their individual effects are hard to distinguish. To further investigate these correlations, Chi-square tests between the pairs of NPIs were performed (See Table 7 for all significant effects) which showed the same significant effects. Based on this analysis more NPIs seem to be related to the closure of the businesses/school closure cluster, like economic measures, strengthening the public health system and lockdown measures. The corresponding heatmap for all the pairwise comparisons for the Apple data is presented in Figure 5, and the significant correlations found for the NPIs used in the Google analysis are presented in Table 8.

**Figure 4. Tree diagram from the cluster analysis showing correlations between NPIs**

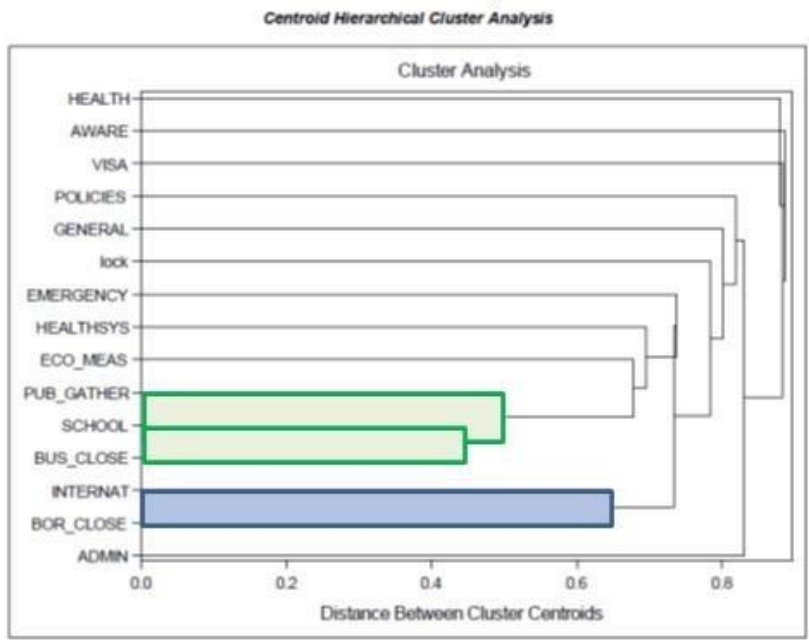

**Table 7. Chi-square associations between NPIs for Apple data**

| Association | P-value |
|---|---|
| Border closure vs International flights suspension | <0.001 |
| Schools closure vs Closure of businesses and public services | <0.001 |
| Strengthening the public health system vs Economic measures | 0.001 |
| Limit public gatherings vs Strengthening the public health system | 0.006 |
| Strengthening the public health system vs Closure of businesses and public services | 0.008 |
| Closure of businesses and public services vs Economic measures | 0.010 |
| Limit public gatherings vs No lockdown, Partial lockdown or Full lockdown | 0.014 |
| Limit public gatherings vs Closure of businesses and public services | 0.020 |
| Emergency administrative structures activated or established vs Economic measures | 0.027 |
| Awareness campaigns vs No lockdown, Partial lockdown or Full lockdown | 0.030 |
| General recommendations vs State of emergency declared | 0.032 |
| No lockdown, Partial lockdown or Full lockdown vs Economic measures | 0.033 |
| Health screenings in airports and border crossings vs Strengthening the public health system | 0.036 |

**Figure 5. P-values for all pairwise comparisons from the Chi-square associations of NPIs for the Apple data**

| NPIs | Emergency administrative structures activated or established | Awareness campaigns | Border closure | Closure of businesses and public services | Economic measures | State of emergency declared | General recommen-dations | Health screenings in airports and border crossings | Strengthening the public health system | International flights suspension | Isolation and quarantine policies | Limit public gatherings | Schools closure | Visa restrictions | Partial/Full Lockdown |
|---|---|---|---|---|---|---|---|---|---|---|---|---|---|---|---|
| Emergency administrative structures activated or established | | 0.227 | 0.716 | 0.917 | 0.027 | 0.200 | 0.991 | 0.519 | 0.145 | 0.417 | 0.665 | 0.889 | 0.889 | 0.273 | 0.070 |
| Awareness campaigns | 0.227 | | 0.803 | 0.471 | 0.388 | 0.508 | 0.530 | 0.908 | 0.145 | 0.869 | 0.849 | 0.583 | 0.440 | 0.273 | 0.030 |
| Border closure | 0.716 | 0.803 | | 0.957 | 0.477 | 0.176 | 0.448 | 0.103 | 0.933 | 0.000 | 0.993 | 0.172 | 0.564 | 0.200 | 0.317 |
| Closure of businesses and public services | 0.917 | 0.471 | 0.957 | | 0.010 | 0.589 | 0.889 | 0.851 | 0.008 | 0.918 | 0.447 | 0.020 | 0.000 | 0.403 | 0.595 |
| Economic measures | 0.027 | 0.388 | 0.477 | 0.010 | | 0.767 | 0.448 | 0.106 | 0.001 | 0.525 | 0.291 | 0.051 | 0.971 | 0.263 | 0.033 |
| State of emergency declared | 0.200 | 0.508 | 0.176 | 0.589 | 0.767 | | 0.032 | 0.210 | 0.396 | 0.770 | 0.326 | 0.100 | 0.206 | 0.697 | 0.831 |
| General recommendations | 0.991 | 0.530 | 0.448 | 0.889 | 0.448 | 0.032 | | 0.855 | 0.098 | 0.946 | 0.159 | 0.300 | 0.511 | 0.832 | 0.726 |
| Health screenings in airports and border crossings | 0.519 | 0.908 | 0.103 | 0.851 | 0.106 | 0.210 | 0.855 | | 0.036 | 0.187 | 0.461 | 0.853 | 0.098 | 0.148 | 0.347 |
| Strengthening the public health system | 0.145 | 0.145 | 0.933 | 0.008 | 0.001 | 0.396 | 0.098 | 0.036 | | 0.651 | 0.939 | 0.006 | 0.795 | 0.957 | 0.592 |
| International flights suspension | 0.417 | 0.869 | 0.000 | 0.918 | 0.525 | 0.770 | 0.946 | 0.187 | 0.651 | | 0.525 | 0.793 | 0.162 | 0.066 | 0.592 |
| Isolation and quarantine policies | 0.665 | 0.849 | 0.993 | 0.447 | 0.291 | 0.326 | 0.159 | 0.461 | 0.939 | 0.525 | | 0.137 | 0.978 | 0.503 | 0.447 |
| Limit public gatherings | 0.889 | 0.583 | 0.172 | 0.020 | 0.051 | 0.100 | 0.300 | 0.853 | 0.006 | 0.793 | 0.137 | | 0.066 | 0.473 | 0.014 |
| Schools closure | 0.889 | 0.440 | 0.564 | 0.000 | 0.971 | 0.206 | 0.511 | 0.098 | 0.795 | 0.162 | 0.978 | 0.066 | | 0.384 | 0.677 |
| Visa restrictions | 0.273 | 0.273 | 0.200 | 0.403 | 0.263 | 0.697 | 0.832 | 0.148 | 0.957 | 0.066 | 0.503 | 0.473 | 0.384 | | 0.783 |
| Partial/Full Lockdown | 0.070 | 0.030 | 0.317 | 0.595 | 0.033 | 0.831 | 0.726 | 0.347 | 0.592 | 0.592 | 0.447 | 0.014 | 0.677 | 0.783 | |

**Table 8. Chi-square associations between NPIs for Google data**

| Association | P-value |
|---|---|
| Schools closure vs Closure of businesses and public services | <0.001 |
| Strengthening the public health system vs Economic measures | <0.001 |
| Border closure vs International flights suspension | 0.001 |
| Closure of businesses and public services vs Economic measures | 0.003 |
| Schools closure vs Curfews | 0.004 |
| Border closure vs Limit public gatherings | 0.005 |
| Strengthening the public health system vs Closure of businesses and public services | 0.005 |
| Limit public gatherings vs No lockdown, Partial lockdown or Full lockdown | 0.009 |
| Limit public gatherings vs Schools closure | 0.020 |
| Limit public gatherings vs Closure of businesses and public services | 0.024 |
| Domestic travel restrictions vs No lockdown, Partial lockdown or Full lockdown | 0.025 |
| Limit public gatherings vs Economic measures | 0.028 |
| Awareness campaigns vs Strengthening the public health system | 0.030 |
| Domestic travel restrictions vs International flights suspension | 0.037 |
| General recommendations vs State of emergency declared | 0.043 |
| Health screenings in airports and border crossings vs Strengthening the public health system | 0.046 |

The results obtained from the selected best fit model from the generalized linear model, analysing both the magnitude of the effect and the rate of the change for the Apple data, are presented in Table 9. In general the NPIs had an effect of 50.4% loss in mobility which was not attributable to a single NPI. Additional significant effects were observed for lockdown measures (13.2%) and the declaration of a state of emergency (7.3%). Awareness campaigns and school closures did seem to have an effect as well. The basic speed of the effect was a reduction of mobility of 4.9% per day. The speed of the effect was significantly increased by 1.8% per day for the lockdown measures, 1.2% per day for limitations of public gatherings and to a lesser extent by school closures (1.0% per day) and the declaration of a state of emergency (0.9% per day).

**Table 9. Estimates and p-values from the selected best fit model from the generalized linear model analysis for the Apple data**

| Endpoint | Parameter | Estimate | P-value |
|---|---|---|---|
| Average % mobility lost (extent of the effect) | Intercept | 50.4 | <0.001 |
| Overall model p-value = < 0.001 | Awareness campaigns | 5.598 | 0.090 |
|  | No lockdown, Partial lockdown or Full lockdown | 13.249 | <0.001 |
|  | School closures | 4.766 | 0.084 |
|  | Strengthening the public health system | -3.917 | 0.167 |
|  | State of emergency declared | 7.261 | 0.008 |
| Slope (speed of the effect) |  |  |  |
| Overall model p-value = <0.001 | Intercept | -4.876 | <0.001 |
|  | Limit public gatherings | -1.222 | 0.075 |
|  | No lockdown, Partial lockdown or Full lockdown | -1.787 | 0.006 |
|  | School closures | -1.028 | 0.080 |
|  | State of emergency declared | -0.93 | 0.101 |

The estimates and p-values from the selected best fit model for the Google RAR and Google TS data are presented in Table 10. As seen with the Apple data, the presence of either a partial or full lockdown in the countries also contributed to the largest increase in the average % mobility lost for Google RAR (16.6%) as well as for Google TS data (16.4%). The declaration of a state of emergency within the countries had a significant increase of 4.9% in the average % mobility lost for the Google TS data, where the closure of businesses and public services had a significant increase of 5.8% in the average % mobility lost for the google RAR data.

Table 10. Estimates and p-values from selected best fit model for the Google data

| Endpoint | Parameter | Estimate | P-value |
|---|---|---|---|
| Google retail and recreation Average % mobility lost (extent of the effect) | Intercept | 45.699 | <0.001 |
| Overall model p-value = < 0.001 | No lockdown, Partial lockdown or Full lockdown | 16.622 | <0.001 |
|  | Closure of businesses and public services | 5.791 | 0.040 |
|  | State of emergency declared | 5.040 | 0.071 |
| Google transit and stations Average % mobility lost (extend of the effect) |  |  |  |
| Overall model p-value = <0.001 | Intercept | 50.344 | <0.001 |
|  | General recommendations | -4.000 | 0.1156 |
|  | No lockdown, Partial lockdown or Full lockdown | 16.371 | <0.001 |
|  | State of emergency declared | 4.868 | 0.046 |

# Discussion

In this study, both the magnitude of the change in mobility and the rate of transition between the pre-lockdown and lockdown status during the COVID-19 pandemic were assessed, using Google and Apple mobility data [6,7]. An average decrease in mobility of 60.6% was observed for the Apple mobility data. The Google RAR data and the Google TS data showed similar, but slightly lower results. The number of unique NPIs implemented within the last 3 weeks before stable low mobility had a significant effect on the % remaining mobility. This was also observed by Islam et al. [9] who saw a combined effect on incidence rates of COVID-19 of 5 of the main NPIs issued. Our finding of close correlation between NPIs might explain why, in their study, the sequence of NPIs did not show a consistent association pattern with COVID-19 incidence. We found that effects of individual NPIs on the decrease in mobility were hard to distinguish as they were often issued within a few days from each other. Two of the most closely associated NPIs were school closures and closure of businesses and public services.

While for Apple data the school closures were modelled as an effect, the Google retail and recreation model selected the closure of businesses and public services. However, due to the close relation, both effects are mixed and cannot clearly be distinguished. Taking that into account, the independent modelling results of both Google models and the Apple model gave corresponding results.

The Apple data showed a 50.4% of mobility loss that was not attributable to a single NPI in the best fit model. The Google TS model was equal in this regard (50.3%), while the Google RAR model was more specific with 45.7% unattributable loss in mobility. All models showed an additional significant reduction in mobility attributable to lockdown measures (full or partial) of 13.2% for the Apple data and 16.6% and 16.4% for the Google RAR and Google TS data respectively. The NPI "state of emergency declared" was selected in all models and, significant ($p<0.05$) in two of them, indicated that this effect was additional to the general lockdown measures. The effect attributable to this NPI ranged on average from a reduction in mobility of 7.2% for the Apple data to a reduction of mobility of 4.9% for the Google TS data. Closure of businesses and public services had a significant effect on the reduction of mobility for the Google RAR data. For the Apple data, the closely correlated NPI "School closures" was selected (effect -4.8%, $p=0.084$). This was to be expected, since businesses and public services would have a direct effect on retail visits and recreation, while school closures would have a more general effect on reduction of mobility in a population due to childcare requirements. In addition, awareness campaigns also seem to have an effect on the mobility (effect -6.0%, $p=0.090$), which indicates that governments educating their population on the impacts and effects of COVID-19 would be more effective in their lockdown measures.

The speed of the effect (the gradient) of the NPIs on mobility was analysed only for the Apple data, due to a non-normal distribution for the corresponding Google data. For this Apple data we observed on average a reduction in mobility of 4.9% per day. Although the extent of lockdown measures probably differed greatly between countries, an average increase in the speed of 1.8% per day was still observed due to the lockdown measures. In addition, limiting public gatherings had an additional significant increase in speed of 1.2% per day. Surprisingly the limiting of public gatherings was selected as having an effect in this model ($p=0.075$), suggesting that it had an additional, more immediate effect on the lockdown measures issued in a country.

Most studies up to the writing of this article assessed the impact of NPIs on COVID-19 case numbers, deaths and the basic reproduction number ($R_0$) [3,5]. Cases, deaths and $R_0$ are difficult to measure, especially in a timely manner; additionally they will largely not include asymptomatic numbers. Mobility data, on the other hand, is widely available, implemented continuously, passively and uniformly in all countries (with the caveat of mobile network penetration and device ownership excepted). Even more accurate behavioural changes in activity after the implementation of NPIs can be observed in greater detail by analysing data collected through smartphones and wearables [10]. As this actively requires the permission and cooperation of the targeted subjects, as well as the analysis of huge amounts of data, this is more feasible for sub-populations in selected countries. The Google and Apple data are anonymous and can therefore be used in more general cases.

As can be seen from our results, the generalized NPIs - as obtained from the ACAPS database - are catalogued with best effort and are probably not consistent in how they are implemented and adhered to in different countries. The strictness in how they are issued will differ from country to country. They are often issued jointly and different NPIs are more or less correlated. The observed effects of NPIs in this study are the most dominant ones. Due to the great variability between countries, additional effects of other NPIs could not be distinguished. To be able to make a clear distinction between countries on how the measures affected COVID-19 disease rates, the % remaining mobility in a country - in combination with the period for which this mobility is low - is probably a better estimate to correlate with the disease activity than the specific NPIs. This was also substantiated by Hale et al. [11] who proposed an aggregate stringency index based on all the NPIs issued in a country. As this index is a combination of different NPIs and their characteristics, it is a better predictor of death rates than the individual NPIs. The variability of NPI implementation and adherence are not provided in the NPI database. As a result it will miss the between-country-effects of some of the more marginal NPIs. The mobility rates which result from issuing the NPIs based on the available Google and Apple data are probably better in reflecting the true effect of NPIs on COVID-19 spread and deaths, as mobility is more directly related. This is illustrated by Drake et al. [4] who examined the effect of NPIs on mobility in the UK using the same Google data. They observed a decrease in adherence to movement restrictions over time. The effects of loosening the NPIs and the reduction in adherence of the population to the NPIs can be better examined using the effect of the NPIs on the mobility rates than the delayed effects on the Covid-19 measures. When assessing the direct effect of NPIs on COVID-19 spread, there will be a lag of at least 14 days due to incubation timing and a delay of disease transfers. Therefore, assessing the effect of implementing and releasing the NPIs in a single country or area on the corresponding mobility will give a good pre-indication of the effectiveness of those measures on the COVID-19 spread.

## Strengths and limitations of this study

Instead of a standardized statistical modelling approach, this model is based on the actual characteristics of the Google and Apple anonymized mobility data during the COVID-19 lockdown period. An accurate measure of % remaining mobility was defined and was applicable for both Apple and Google data. Visual inspection of the results showed an accurate overlay of the estimated % remaining mobility and the actual data. In a number of countries, the mobility percentage was less stable. Hence, the increase of activity started again after a few days of the stabilization. Although the resulting estimated % remaining mobility is in these cases higher than the lowest mobility reported, it still gives a good representation of the mobility in the first two weeks of lockdown.

Both Apple and Google data, as is typical of mobility data, showed strong weekday periodic effects which varied considerably from country to country, during different periods and also in magnitude. The weekday periodic effects were generally more pronounced before stabilization and degraded thereafter. Although Google had corrected their data for average effects on weekdays, the weekday effects were still apparent in many cases. This is probably due to over- or under-compensation in the different periods of the whole observation period. Some countries showed weekday effects before the stabilization, some countries showed weekday effects after the stabilization and other countries showed weekday effects during the complete observation period. For the estimation of the % remaining mobility, the weekday effect was reduced in the period after the stabilization by subtracting the average weekday deviation from the mobility. However, due to inconsistencies in the effects, there was still quite a lot of variation in the data after correcting for weekday effects, both in the pre-lockdown period as during the lockdown period. Therefore it was not possible to assess the start of decline and start of stability using standardized methods like deriving inflexion points from Loess of fit data.

In some cases significant events around the stabilization, other than NPIs, had an effect on the mobility - like the flooding in Egypt of 11 March 2020 [12], Independence Day of 24 February 2020 in Estonia and the elections of 29 February 2020 in Slovakia [Figure 6]. Upon viewing the resulting parameters, these events did not seem to have a significant effect on the % remaining mobility or the speed of the decline of mobility. In some cases it had an effect on the start of the decline date. However, it did not affect the gradient of the decline.

**Figure 6. Significant events in Egypt and Estonia**

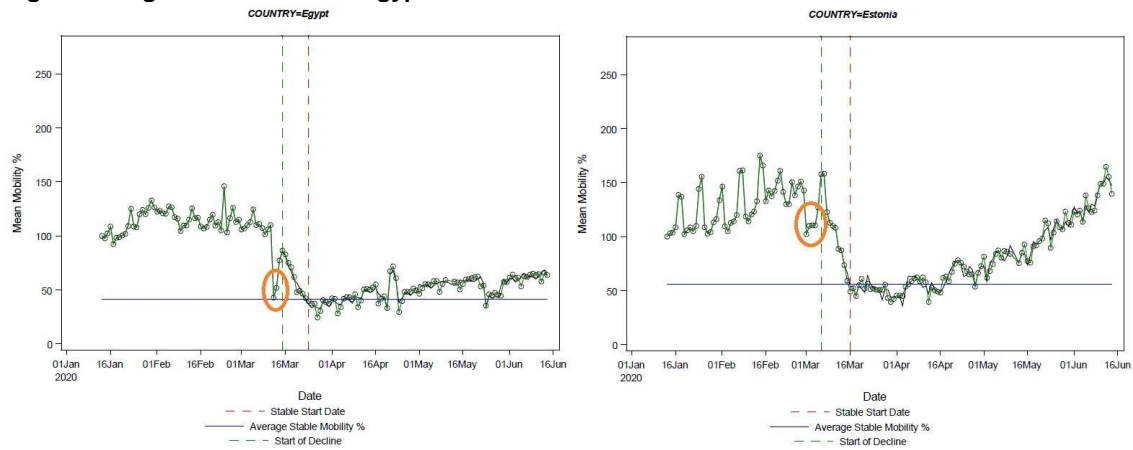

Due to countries' sudden awareness for the need to constrain mobility, the NPIs were often issued at the same time or within a few days from each other. Effects for an individual NPI are therefore more difficult to distinguish. The cluster analysis indicates that NPIs, like school closures and closure of business and public services, were often issued at the same time in the same country. This also holds for international flight suspensions and border closures. For the chosen model, all NPIs were included as separate effects and the NPIs that had the strongest effect on the % mobility lost were selected. However, as the NPIs are clustered, these effects should be viewed in relation to the NPIs that occurred collaterally.

The nature of the Apple data, Google TS data and that of the Google RAR data differed. The countries for which the mobility data was available differed as well. Although Google data was available for more countries than Apple data, there were also countries in which Apple data was available and Google data not. The extent to which the available Apple and Google mobility data reflect the true behaviour of the countries' population was affected by the accessibility to and usability of smartphones. For the Apple mobility data, only those movements for which directions were requested have been reflected in the data. The representativeness of the mobility data further depended on the nature of the population, their inclination to follow the NPI rules and potential data restrictions of the government. Despite that, as the results of the different generalized linear model analyses and the mean decrease in % mobility were comparable across the Apple and Google data, there is a strong indication that, in general, estimates obtained from these models are consistent and indicative of the actual effects on mobility. A combination of different mobility measures like the Apple data, the Google TS data and the Google RAR data are very good and readily available metrics for assessment of the impact of the NPIs issued in the countries. We see the utilization of this data not only for assessing NPI efficacy, but also to monitor local outbreaks and subsequent control measures as countries emerge from nationally implemented NPIs.

## Competing interests

The authors declare that there are no competing interests.

## Author contribution

A.F. and B.S. conceived of the presented idea and B.S. developed the theory. M.B. collected the data, performed the computations and contributed to the methods and results section of the manuscript. B.S. verified the analytical methods and wrote the manuscript. A.F., R.D. and S.S. provided critical feedback and helped shape the research, analysis and manuscript.